\documentclass[a4paper]{article}

\usepackage{INTERSPEECH2021}
\usepackage{multirow}
\usepackage{makecell}
\usepackage{pifont}
\usepackage{amssymb}
\usepackage{graphicx}
\usepackage{subfigure}

\usepackage{array}
\usepackage{float}
\usepackage{cite}
\usepackage[hang,flushmargin]{footmisc} 

\newcommand{\cmark}{\ding{51}}
\newcommand{\xmark}{\ding{55}}
\newcommand{\tabincell}[2]{\begin{tabular}{@{}#1@{}}#2\end{tabular}}

\title{Spectro-Temporal Deep Features \\ for Disordered Speech Assessment and Recognition}
\name{Mengzhe Geng$^1$, Shansong Liu$^1$, Jianwei Yu$^1$, Xurong Xie$^{2}$, Shoukang Hu$^1$, Zi Ye$^1$, Zengrui Jin$^1$, Xunying Liu$^1$, Helen Meng$^1$}
\address{
  $^1$The Chinese University of Hong Kong\\
  $^2$Shenzhen Institutes of Advanced Technology, Chinese Academy of Sciences}
\email{\{mzgeng,ssliu,jwyu,skhu,zye,zrjin,xyliu,hmmeng\}@se.cuhk.edu.hk, xr.xie@siat.ac.cn}

\begin{document}

\maketitle
\begin{abstract}
Automatic recognition of disordered speech remains a highly challenging task to date. Sources of variability commonly found in normal speech including accent, age or gender, when further compounded with the underlying causes of speech impairment and varying severity levels, create large diversity among speakers. To this end, speaker adaptation techniques play a vital role in current speech recognition systems. Motivated by the spectro-temporal level differences between disordered and normal speech that systematically manifest in articulatory imprecision, decreased volume and clarity, slower speaking rates and increased dysfluencies, novel spectro-temporal subspace basis embedding deep features derived by SVD decomposition of speech spectrum are proposed to facilitate both accurate speech intelligibility assessment and auxiliary feature based speaker adaptation of state-of-the-art hybrid DNN and end-to-end disordered speech recognition systems. Experiments conducted on the UASpeech corpus suggest the proposed spectro-temporal deep feature adapted systems consistently outperformed baseline i-Vector adaptation by up to 2.63\% absolute (8.6\% relative) reduction in word error rate (WER) with or without data augmentation. Learning hidden unit contribution (LHUC) based speaker adaptation was further applied. The final speaker adapted system using the proposed spectral basis embedding features gave an overall WER of 25.6\% on the UASpeech test set of 16 dysarthric speakers. 
\end{abstract}
\noindent\textbf{Index Terms}: Speech Disorders, Speech Recognition, Speaker Adaptation, Speech Assessment, Subspace-based Learning

\section{Introduction}

In spite of the swift progress of automatic speech recognition (ASR) technologies targeting normal speech in the past few decades\cite{bahl1986maximum, povey2002minimum, graves2006connectionist,graves2013speech, peddinti2015time, chan2016listen,wang2020transformer,hu2021neural,hu2021bayesian}, accurate recognition of disordered speech remains a demanding task to date \cite{christensen2012comparative, christensen2013combining,sehgal2015model,yu2018development,hu2019cuhk,liu2020exploiting,wang2021improved}. The underlying causes of speech disorders include a wide range of neuro-motor conditions, such as cerebral palsy, Parkinson disease, amyotrophic lateral sclerosis and stroke or traumatic brain injuries \cite{lanier2010speech}. Despite the reduced intelligibility of disordered speech, hands-free and speech-enabled assistive technologies are among natural alternatives \cite{hawley2005speech} to aid people with speech disorders since they often suffer from the co-occurring physical disabilities.

A key challenge in current ASR system development is to systematically model the latent variations among diverse speech data. A wide range of sources of variability commonly found in normal speech, including speaker specific idiosyncrasy such as accent and physiological differences brought by age or gender, when further compounded with the underlying inducements of speech impairment and varying severity levels, create large diversity among disordered speech. Due to the associate difficulties in controlling the muscles and articulators used in speech production \cite{hixon1964restricted}, abnormalities including articulation imprecision, reduced intensity and clarity, slower speaking rates and increased disfluencies are observed in disordered speech \cite{kent2000dysarthrias}. Furthermore, temporal or spectral perturbation based data augmentation techniques widely used in both state-of-the-art ASR systems for normal speech \cite{jaitly2013vocal, kanda2013elastic, cui2015data, ko2015audio} and recently those designed for impaired speech \cite{vachhani2018data, xiong2019phonetic, geng2020investigation} introduce extra diversity. To this end, speaker adaptation techniques play a crucial role in current ASR systems for both normal and disordered speech.

Previous researches of speech adaptation targeting normal speech recognition can be divided into three broad categories: 1) auxiliary speaker embedding feature based approaches that encodes speaker-dependent (SD) characteristics in a compact vector representation, such as speaker codes \cite{abdel2013fast}, i-Vectors \cite{saon2013speaker} and bottleneck features \cite{huang2015investigation};  2) feature transformation based methods applied at acoustic front-ends that produce speaker invariant input features in a canonical representation, such as feature-space maximum likelihood linear regression (f-MLLR) \cite{gales1998maximum}; 3) model based adaptation techniques that exploits specially designed SD transformations in model parameters to handle the speaker level variability \cite{anastasakos1996compact, swietojanski2016learning}.

In contrast, so far there has been limited research on speaker adaptation targeting disordered speech recognition, particularly those suitable for state-of-the-art ASR systems. Many of the earlier researches were conducted in the context of traditional hidden Markov models (HMMs) with Gaussian mixture model (GMM) state density distributions. In \cite{mengistu2011adapting, christensen2012comparative, kim2013dysarthric}, maximum likelihood linear regression (MLLR) and maximum a posterior (MAP) were applied to speaker-independent (SI) GMM-HMM systems. In \cite{sehgal2015model}, a combination of MLLR and MAP adaptation were used in speaker adaptive training (SAT) of SI GMM-HMM models. In \cite{bhat2016recognition}, f-MLLR based SAT was studied. In \cite{kim2017regularized}, regularized speaker adaptation on Kullback-Leibler divergence-based HMMs (KL-HMMs) was conducted. More recent researches investigated model adaptation of state-of-the-art deep neural network (DNN) based systems. Dysarthric speaker adaptation of recurrent neural network transducers (RNN-Ts) \cite{graves2013speech} and lattice-free MMI trained time delay neural networks (TDNNs) \cite{peddinti2015time} via direct model parameter fine-tuning were studied in \cite{shor2019personalizing, xiong2020source}. Learning hidden unit contributions based (LHUC) SAT \cite{swietojanski2016learning} was investigated in  \cite{yu2018development, geng2020investigation}. The majority of prior researches on disordered speech adaptation focused on feature transformation and model based adaptation. On the contrary, very limited research has been conducted on auxiliary speaker embedding feature based adaptation approaches, particularly those motivated by the underlying spectra-temporal variation of impaired speech of diverse causes and severity levels.

This paper proposes novel deep spectro-temporal subspace basis embedding features to facilitate both accurate speech intelligibility assessment and auxiliary feature based speaker adaptation for disordered speech recognition. Spectral and temporal basis vectors derived by singular value decomposition (SVD) \cite{van1993subspace} of speech spectrum were used to structurally represent the spectro-temporal level key attributes found in disordered speech \cite{rosen2006parametric, kacha2020principal,janbakhshi2020subspace}, such as the overall decrease in speaking rate and speech volume as well as changes in spectral envelope. These two form of basis vectors were then used to construct a DNN speech intelligibility classifier. More compact, lower dimensional speaker specific spectral and temporal DNN embedding features produced by the bottleneck layer of the resulting intelligibility classifier were further employed as auxiliary features to adapt start-of-the-art hybrid DNN \cite{geng2020investigation} and CTC end-to-end \cite{graves2006connectionist} disordered speech recognition systems.  Experiments were conducted on the largest available and most widely used UASpeech \cite{kim2008dysarthric} dysarthric speech corpus and LHUC based speaker adaptation was further applied.

The main contributions of the paper are summarized below:

1) To the best of our knowledge, our novel spectro-temporal deep feature based adaptation approach is the first work to exploit auxiliary speaker embedding features in disordered speech adaptation. In contrast, prior works \cite{mengistu2011adapting, christensen2012comparative, kim2013dysarthric,sehgal2015model,bhat2016recognition,kim2017regularized,yu2018development,geng2020investigation,xiong2020source,shor2019personalizing} focused on feature transformation and model based adaptation. Speaker embedding features, e.g. i-Vector \cite{an2015automatic,garcia2018multimodal}, were used in speech assessment rather than ASR adaptation tasks.


2) The spectro-temporal deep features are intuitively related to the underlying diversity of disordered speech. The spectral basis embedding features are designed to learn characteristics such as volume reduction, changes of formant position, imprecise articulation and hoarse voice while the temporal ones to capture patterns such as increased disfluencies and pauses. Experiments conducted on UASpeech suggest that our proposed spectro-temporal deep feature adapted systems consistently outperformed comparable baseline i-Vector adaptation \cite{saon2013speaker, senior2014improving} by up to 2.63\% absolute (8.6\% relative) reduction in word error rate (WER) with or without data augmentation. The final speaker adapted system using the proposed spectral basis embedding features gave an overall WER of 25.6\% on the UASpeech test set of 16 dysarthric speakers, which is the best ASR performance so far published as far as we know.

The rest of this paper is organized as follows. The derivation of spectro-temporal basis vectors via speech spectrum subspace decomposition is presented in Sec.2. The subspace basis vector based DNN speech intelligibility classifier and embedding features for speaker adaptation are proposed in Sec.3. Sec.4 presents experiments and results of both speech intelligibility assessment and speech recognition on UASpeech. The last section concludes and discuss possible future works.

\section{Speech Spectrum Subspace Decomposition}

To systematically reveal the patterns contained in disordered speech, we conduct SVD on the mel-filterbank log amplitude spectrum following \cite{van1993subspace} to derive basis vectors of the spectral and the temporal subspaces. Let $\mathbf{S_r}$ represent a $C \times T$ dimensional mel-spectrogram of utterance $r$ with $C$ mel-filterbank channels and $T$ frames. The SVD of $\mathbf{S_r}$ is given by

\begin{equation}
\setlength{\abovedisplayskip}{3pt}
\setlength{\belowdisplayskip}{3pt}
\mathbf{S_{r} = U_{r}{\Sigma_{r}}V_{r}^{\mathrm{T}}}
\end{equation}

where the set of column vectors of the $C\times C$ dimensional $\mathbf{U_{r}}$ matrix (the left-singular vectors) and the set of row vectors of the $T \times T$ dimensional $\mathbf{V_{r}^\mathrm{T}}$ matrix (the right-singular vectors) are respectively the bases of the spectral and the temporal subspaces, and $\mathbf{\Sigma_{r}}$ is a $C \times T$ diagonal matrix containing the singular values in descending order \cite{van1993subspace}. The rank of $\mathbf{S_{r}}$ is equal to the number of non-zero singular values, i.e. $rank(\mathbf{S_{r}}) \leq min\{C,T\}$. Motivated by low-rank approximation \cite{fevotte2009nonnegative}, we select the top d principal spectral and temporal basis vectors for all the experiments of this paper.

\begin{figure}[ht]
  \centering
  \vspace{-0.2cm} 
  \setlength{\abovecaptionskip}{0.2cm}   
  \setlength{\belowcaptionskip}{-0.35cm}   
  \includegraphics[scale=0.13]{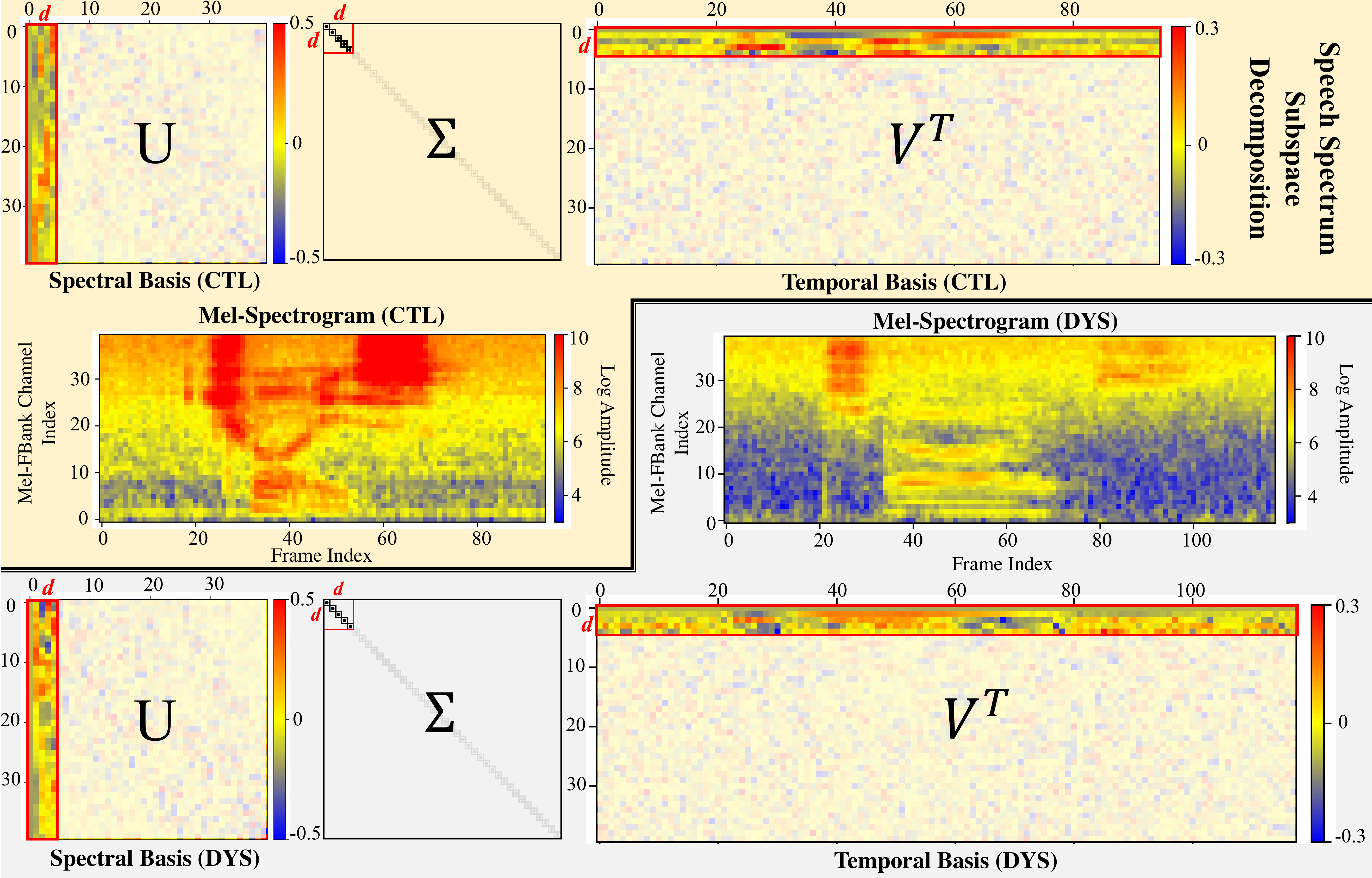}
  \caption{Example subspace decomposition of mel-spectrograms of normal (CTL, upper) and dysarthric (DYS, lower) utterances of word ``choice'' to obtain top $d$ spectral and temporal basis vectors (circled in red in $\mathbf{U}$ and $\mathbf{V^\mathrm{T}}$). }
  \label{fig:SVD-Example}
\end{figure}  

As shown in Fig.\ref{fig:SVD-Example}, when compared with that of the normal speaker, the mel-spectrogram of the dysarthric speaker contains reduced energy especially in the lower portions of mel-scale frequencies. The overall spectral envelope of the dysarthric spectrum is less visible with weakened formants. Spectral basis vectors are designed to capture these patterns as well as imprecise articulation and hoarse voice. As the dysarthric speaker speaks more slowly and less fluently, the extracted temporal basis vectors encode the speaking rate in their dimensionality in addition to other patterns such as increased disfluencies and pauses.

\begin{figure}[ht]
  \centering
  \vspace{-0.2cm} 
  \setlength{\abovecaptionskip}{0.2cm}   
  \setlength{\belowcaptionskip}{-0.6cm}   
  \includegraphics[scale=0.3]{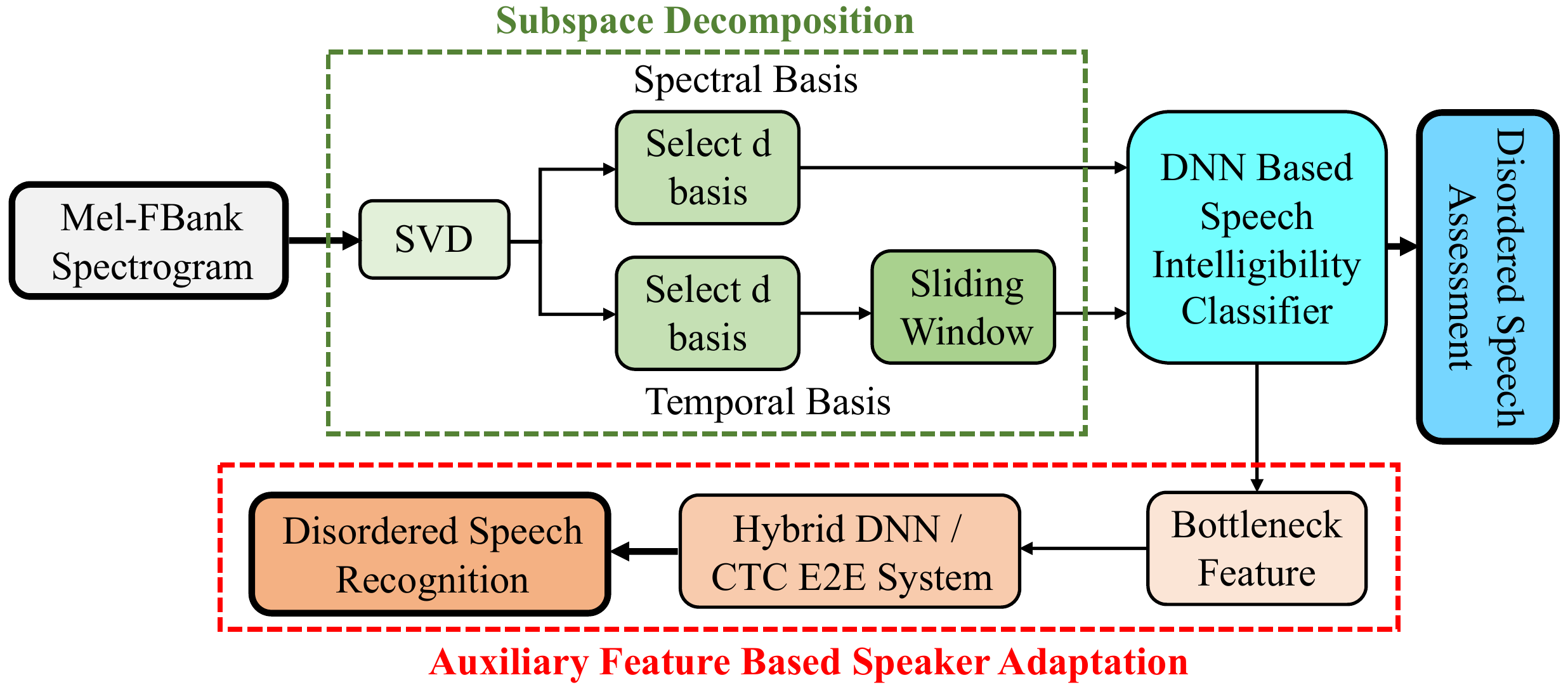}
  \caption{Framework of the proposed speech intelligibility assessment and auxiliary feature based speaker adaptation founded on spectro-temporal subspace basis vectors derived from speech spectrum subspace decomposition. }
  \label{fig:SVD-procedure}
\end{figure}

Traditional disordered speech assessment methods often require the contents spoken by different speakers to be the same \cite{bocklet2013automatic,janbakhshi2020subspace}, which allows normal and dysarthric speech of identical contents but varying durations to be compared after an alignment procedure \cite{berndt1994using}. In order to facilitate a more practical assessment scheme applicable to unrestricted speech contents of unknown durations and derive speaker-level embedding vectors of consistent dimensionality for ASR system adaptation in this paper, when processing the temporal basis vectors of each utterance, a frame-level sliding window of 25 dimensions was applied to the top $d$ selected temporal basis vectors.  Their 25 dimensional mean and standard deviation vectors were then computed to serve as the``average'' temporal basis representations of fixed dimensionality, as shown in Fig.\ref{fig:SVD-procedure}.

\section{Spectro-Temporal Deep Features}

The overall framework of our proposed subspace decomposition based speech intelligibility assessment and spectro-temporal deep feature based speaker adaptation for disordered speech recognition is shown in Fig.\ref{fig:SVD-procedure}. A DNN speech intelligibility classifier is trained using the spectro-temporal basis vectors discussed in Sec.2. More compact, lower dimensional features obtained from the DNN classifier are used as auxiliary speaker embedding features for disordered speech adaptation.

\subsection{DNN Speech Intelligibility Classifier}

\begin{figure}[ht]
  \centering
  \vspace{-0.2cm} 
  \setlength{\abovecaptionskip}{0.2cm}   
  \setlength{\belowcaptionskip}{-0.5cm}   
  \includegraphics[scale=0.5]{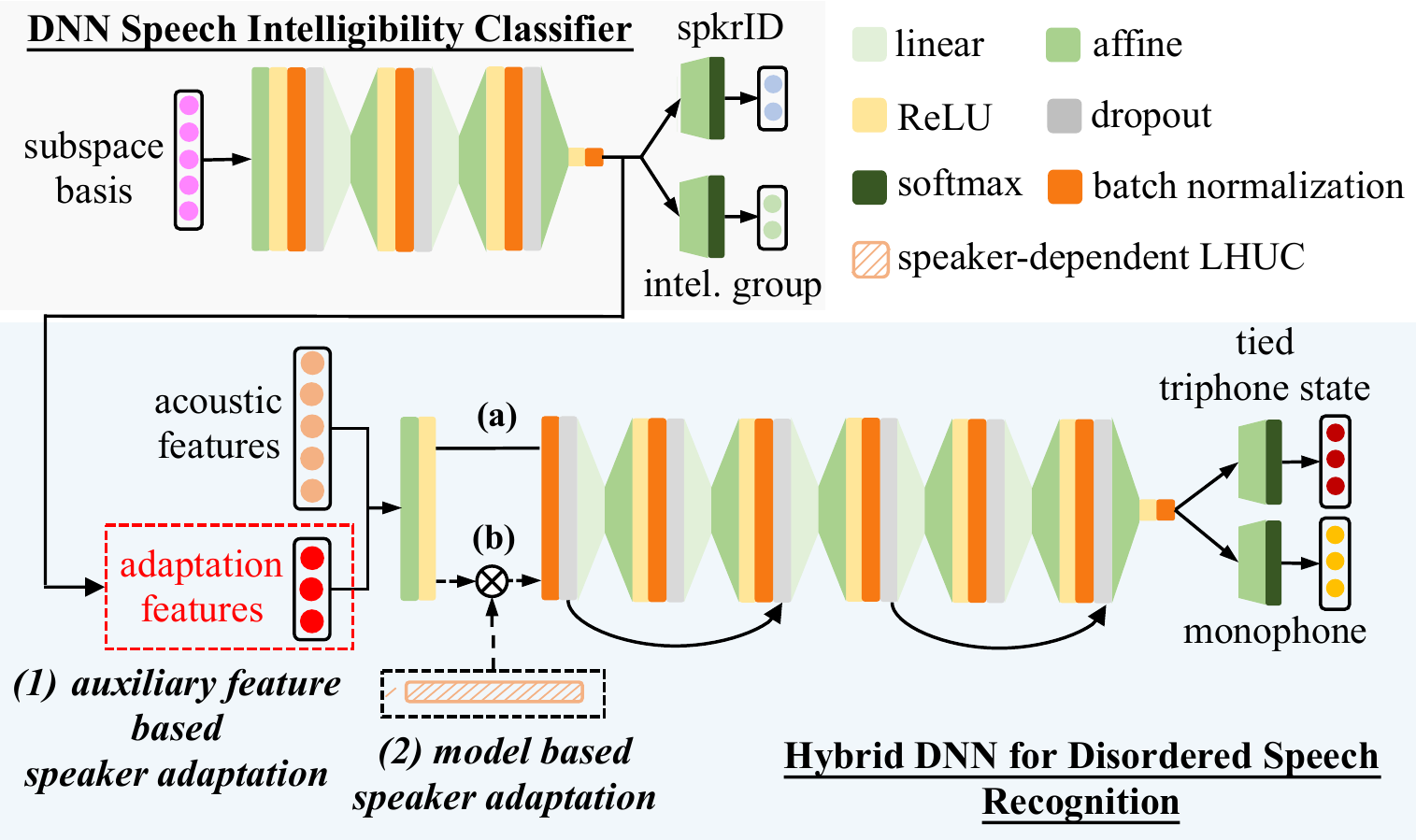}
  \caption{Architecture of the DNN speech intelligibility classifier (upper) and the hybrid-DNN system for disordered speech recognition (lower). Selecting connection (a) leads to systems using auxiliary feature based adaptation only, while selecting connection (b) leads to systems with additional LHUC SAT.}
  \label{fig:hybrid-DNN}
\end{figure}

The DNN speech intelligibility classifier shares similar structure with the hybrid DNN acoustic model for disordered speech recognition except that the DNN classifier has four hidden layers while the hybrid DNN has seven. As illustrated in Fig.\ref{fig:hybrid-DNN}, each hidden layer of the DNN classifier contains a basic set of neural operations performed in sequence, i.e. affine transformation (in green), rectified linear unit (ReLU) activation (in yellow) and batch normalization (in orange).  Apart from this, linear bottleneck projections (in light green) are applied to the inputs of the intermediate two hidden layers and dropout operations (in grey) are applied to the outputs of the first three hidden layers. A skip connection connects the output of the first layer to that of the third layer. The first three layers are of 2000 dimensions while the 25-dimensional fourth layer serves at the bottleneck layer. Softmax activation (in dark green) is applied in the output layer. Multi-task learning (MTL) \cite{caruana1997multitask} is implemented and the labels for the two tasks are the speaker's intelligibility group and speaker ID respectively.  


\subsection{Auxiliary Feature Based Speaker Adaptation}


As shown in Fig.\ref{fig:SVD-procedure} and Fig.\ref{fig:hybrid-DNN}, the output of the 25 dimensional bottleneck layer, after a speaker-level averaging, is concatenated to the acoustic features at the front-end of the hybrid DNN (shown in Fig.\ref{fig:hybrid-DNN}) or CTC end-to-end system\footnote{The CTC E2E system consists of four 2D convolutional layers and three BLSTM layers.} for disordered speech recognition. The hybrid DNN acoustic model shares the same structure as our previous work \cite{geng2020investigation} except for the front-end. The MTL labels for speech recognition are tied triphone state and monophone alignments. In Fig.\ref{fig:hybrid-DNN}, \textbf{1)} Excluding both the adaptation features concatenated at input and the speaker dependent LHUC \cite{swietojanski2016learning} model parameters leads to the baseline speaker independent (SI) systems. \textbf{2)} Keeping the concatenated adaptation features but excluding the LHUC parameters leads to the spectro-temporal deep feature adapted systems. \textbf{3)} Keeping both adaptation features LHUC parameters leads to the spectro-temporal deep feature adapted systems with LHUC SAT.

\section{Experiments and Results}

\subsection{Task Description}
The UASpeech \cite{kim2008dysarthric} corpus is the largest available and most widely used corpus for research on speech disorders. It is an isolated word recognition task consisting of speech recordings from 16 dysarthric and 13 control speakers on 155 common words and 300 uncommon words. There're three blocks per speaker, each containing all 155 common words and one third of the uncommon words. Speech intelligibility rating  given to each speaker lies in five groups, i.e. ``very low", ``low", ``mid", ``high" and ``control". Block 1 and block 3 of all 29 speakers are treated as training set while block 2 of the 16 dysarthric speakers as test set. Silence stripping was performed using a HTK \cite{young2006htk} trained GMM-HMM system as described in our previous work \cite{yu2018development}, which results in a 30.6-hour training set (99195 utterances) and a 9-hour test set (26520 utterances).

\subsection{Experiment Setup}
In our experiments, the DNN speech intelligibility classifier and the hybrid DNN acoustic model were implemented using an extension to the Kaldi toolkit \cite{povey2011kaldi} while the CTC end-to-end system was implemented using PyTorch. We select top 2 and top 5 principal spectral and temporal basis vectors based on the experimental results \cite{kacha2020principal}. The inputs to the acoustic models were 80-dimensional filter bank (FBank) + $\Delta$ features plus 25-dimensional DNN bottleneck spectro-temporal features or 100-dimensional i-Vectors\footnote{We follow Kaldi: egs/wsj/s5/steps/nnet/ivector/extract\_ivectors.sh}. The other settings of the hybrid DNN 
were exactly the same as our previous work \cite{geng2020investigation}. Following \cite{christensen2012comparative}, a uniform language model was used in decoding.

\subsection{Performance of Speech Intelligibility Assessment}

\begin{table} [ht]
    \vspace{-5mm}
	\setlength{\abovecaptionskip}{-0.2cm}
    \caption{Utterance-level accuracy of the DNN speech intelligibility classifier on 49717 utterances from block 2 of all 29 speakers. Here `DYS" is short for dysarthric. ``SB'' and ``TB" denote "spectral basis vectors" and "temporal basis vectors" respectively. "CTL v.s. DYS" and "5-way intel." stand for binary speech intelligibility (normal / dysarthric) assessment and finer classification among DYS intelligibility subgroups including very low (VL), low (L), mid (M), high (H) and control (CTL).}
    \label{tab:assessment-result}
    \centering
    \renewcommand\arraystretch{1.0}
    \renewcommand\tabcolsep{2.0pt}

\scalebox{0.9}{\begin{tabular}{c|c|c|cccc|c|c|c}
\multicolumn{1}{c}{}      & \multicolumn{1}{c}{}          & \multicolumn{1}{c}{}            &      &      &      & \multicolumn{1}{c}{} & \multicolumn{1}{c}{} & \multicolumn{1}{c}{} &                       \\
\hline\hline
\multirow{3}{*}{\tabincell{c}{DNN\\Input}} & \multirow{3}{*}{\tabincell{c}{DNN\\Label}}     & \multirow{3}{*}{Assess}         & \multicolumn{7}{c}{Accuracy(\%)}                                                                                \\
\cline{4-10}
                          &                               &                                 & \multicolumn{5}{c|}{DYS}                                         & \multirow{2}{*}{CTL} & \multirow{2}{*}{Avg}  \\
\cline{4-8}
                          &                               &                                 & VL   & L    & M    & H                    & Avg                  &                      &                       \\
\hline\hline
SB                        & \multirow{2}{*}{Intel.}       & \multirow{5}{*}{\tabincell{c}{5-way\\Intel.}}   & 92.8 & 95.6 & 94.5 & 95.8                 & 94.9                 & 99.3                 & 96.9                  \\
SB+TB                     &                               &                                 & 91.1 & 94.9 & 94.4 & 95.5                 & 94.2                 & 99.3                 & 96.6                  \\
\cline{1-2}\cline{4-10}
SB                        & \multirow{3}{*}{\tabincell{c}{Intel.\\+\\spkrID}} &                                 & 93.4 & 96.7 & 96.5 & 96.3                 & 95.9                 & 99.5                 & 97.6                  \\
TB                        &                               &                                 & 78.2 & 69.4 & 61.8 & 58.6                 & 66.2                 & 93.6                 & 79.0                  \\
SB+TB                     &                               &                                 & \textbf{93.9} & 96.8 & 97.2 & 95.7                 & \textbf{95.9}                 & 99.7                 & \textbf{97.7}                  \\
\hline\hline
SB                        & \multirow{2}{*}{Intel.}       & \multirow{5}{*}{\tabincell{c}{DYS\\v.s.\\CTL}} & -    & -    & -    & -                    & 98.9                 & 99.3                 & 99.1                  \\
SB+TB                     &                               &                                 & -    & -    & -    & -                    & 98.8                 & 99.3                 & 99.0                  \\
\cline{1-2}\cline{4-10}
SB                        & \multirow{3}{*}{\tabincell{c}{Intel.\\+\\spkrID}} &                                 & -    & -    & -    & -                    & \textbf{99.3}                 & 99.5                 & \textbf{99.4}                  \\
TB                        &                               &                                 & -    & -    & -    & -                    & 87.2                 & 93.6                 & 90.2                  \\
SB+TB                     &                               &                                 & -    & -    & -    & -                    & 98.9                 & 99.7                 & 99.3                  \\
\hline\hline
\multicolumn{1}{c}{}      & \multicolumn{1}{c}{}          & \multicolumn{1}{c}{}            &      &      &      & \multicolumn{1}{c}{} & \multicolumn{1}{c}{} & \multicolumn{1}{c}{} &
\end{tabular}}
\vspace{-5mm}
\end{table}

For speech intelligibility assessment, apart from block 2 of the 16 dysarthric speakers, block 2 of the 13 control speakers are also used for testing the DNN speech intelligibility classifier, making a total of 49717 utterances. As shown in Table \ref{tab:assessment-result}, for the 5-way intelligibility assessment, the best overall accuracy of 97.7\% is achieved when both spectral and temporal basis vectors are given as input to the DNN classifier (line 5). For the binary assessment which classifies a speaker as control or dysarthric, the best overall accuracy of 99.4\% is achieved with spectral basis as input (line 8). Ablation study which removes speaker ID from the DNN label (line 1-2, 6-7) indicates that there is only small degradation in assessment accuracy if the intelligibility group labels only are provided during training.

\subsection{Performance of Speaker Adaptation}

\begin{table}[ht]
    \vspace{-5mm}
	\setlength{\abovecaptionskip}{-0.2cm}
    \caption{WER performance comparison of the proposed spectro-temporal deep feature adaptation, i-Vector adaptation and LHUC adaptation on the UASpeech test set of 16 dysarthric speakers. ``6M'' and ``26M'' refer to \# of network parameters. ``DYS" and ``CTL" in ``Data Aug" column denote perturbing the disordered and the normal speech respectively for data augmentation. ``SBE'' and ``TBE'' denote spectral basis and temporal basis embedding features. ``SBE$\star$" and ``STBE$\star$'' mean only intelligibility group labels were used in the previous DNN speech intelligibility classifier for ablation study. ``VL / L / M / H" refer to intelligibility groups.``$\dag$ and ``$\ddag$ denote a statistically significant improvement is obtained over the comparable baseline i-Vector or SI systems and LHUC adaptation respectively.}
    \label{tab:recog-result}
    \centering
    \renewcommand\arraystretch{1.0}
    \renewcommand\tabcolsep{2.0pt}
\scalebox{0.78}{\begin{tabular}{c|c|c|c|c|c|cccc|c}
\multicolumn{1}{c}{}  & \multicolumn{1}{c}{}         & \multicolumn{1}{c}{}                   & \multicolumn{1}{c}{}      & \multicolumn{1}{c}{}        & \multicolumn{1}{c}{}                   &          &       &       & \multicolumn{1}{c}{} &          \\
\hline\hline
\multirow{2}{*}{Sys.} & \multirow{2}{*}{\tabincell{c}{Model\\(\# Par.)}}        & \multirow{2}{*}{\tabincell{c}{Data\\Aug.}}             & \multirow{2}{*}{\# Hrs} &\multirow{2}{*}{\tabincell{c}{Adapt.\\Feat.}}  & \multirow{2}{*}{\tabincell{c}{LHUC\\SAT}}             & \multicolumn{5}{c}{WER\%}                                  \\
\cline{7-11}
                      &                               &                                         &                           &                             &                                        & VL    & L     & M     & H                    & Avg    \\
\hline\hline
1                     & \multirow{10}{*}{\tabincell{c}{Hybrid\\DNN\\(6M)}} & \multirow{10}{*}{\xmark} & \multirow{10}{*}{30.6}    & \xmark       & \multirow{7}{*}{\xmark} & 69.82 & 32.61 & 24.53 & 10.40                & 31.45  \\
2                     &                               &                                         &                           & i-Vector                    &                                        & 67.25 & 32.70 & 22.56 & 10.11                & 30.46  \\
3                     &                               &                                         &                           & TBE                         &                                        & 70.87 & 35.06 & 21.11 & 10.84                & 31.81  \\
4                     &                               &                                         &                           & SBE                         &                                        & \textbf{64.43} & 29.71 & 19.84 & 8.57                 & \textbf{28.05$^\dag$}  \\
5                     &                               &                                         &                           & SBE+TBE                     &                                        & 64.54 & 29.13 & 18.90 & 8.69                & \textbf{27.83$^\dag$}  \\
\cline{1-1}\cline{5-5}\cline{7-11}
6                     &                               &                                         &                           & SBE$\star$                        &                                        & 66.23 & 29.71 & 19.29 & 9.07                 & 28.49$^\dag$  \\
7                     &                               &                                         &                           & SBE+TBE$\star$                       &                                        & 65.77 & 30.06 & 20.76 & 8.70                 & 28.64$^\dag$  \\
\cline{1-1}\cline{5-11}
8                     &                               &                                         &                           & \xmark       & \multirow{3}{*}{\cmark} & 64.39 & 29.88 & 20.27 & 8.95                 & 28.29  \\
9                     &                               &                                         &                           & SBE                         &                                        & 63.40 & 28.90 & 18.64 & 8.13                 & \textbf{27.24$^\ddag$}  \\
10                    &                               &                                         &                           & SBE+TBE                     &                                        & \textbf{63.01} & 28.10 & 18.25 & 8.22                 & \textbf{26.90$^\ddag$}  \\
\hline\hline
11                    & \multirow{7}{*}{\tabincell{c}{Hybrid\\DNN\\(6M)}}  & \multirow{7}{*}{DYS}                    & \multirow{7}{*}{65.9}     & \xmark       & \multirow{4}{*}{\xmark} & 68.43 & 29.60 & 21.37 & 10.44                & 29.79  \\
12                    &                               &                                         &                           & i-Vector                    &                                        & 66.06 & 31.16 & 20.27 & 8.86                 & 28.95  \\
13                    &                               &                                         &                           & SBE                         &                                        & 62.56 & 28.33 & 18.21 & 9.01                 & \textbf{27.13$^\dag$}  \\
14                    &                               &                                         &                           & SBE+TBE                     &                                        & \textbf{61.40} & 28.93 & 18.74 & 9.00                 & \textbf{27.14$^\dag$}  \\
\cline{1-1}\cline{5-11}
15                    &                               &                                         &                           & \xmark       & \multirow{3}{*}{\cmark} & 60.99 & 28.20 & 18.86 & 8.41                 & 26.69  \\
16                    &                               &                                         &                           & SBE                         &                                        & 60.98 & 27.29 & 17.96 & 8.54                 & \textbf{26.32}  \\
17                    &                               &                                         &                           & SBE+TBE                     &                                        & \textbf{60.23} & 27.87 & 18.27 & 8.67                 & \textbf{26.41}  \\
\hline\hline
18                    & \multirow{7}{*}{\tabincell{c}{Hybrid\\DNN\\(6M)}}  & \multirow{7}{*}{\tabincell{c}{DYS\\+\\CTL}}          & \multirow{7}{*}{130.1}    & \xmark       & \multirow{4}{*}{\xmark} & 66.45 & 28.95 & 20.37 & 9.62                 & 28.73  \\
19                    &                               &                                         &                           & i-Vector                    &                                        & 65.52 & 30.63 & 19.27 & 8.60                 & 28.42  \\
20                    &                               &                                         &                           & SBE                         &                                        & 61.55 & 27.52 & 17.31 & 8.22                 & \textbf{26.26$^\dag$}  \\
21                    &                               &                                         &                           & SBE+TBE                     &                                        & \textbf{61.24} & 27.77 & 17.45 & 8.31                 & \textbf{26.32$^\dag$}  \\
\cline{1-1}\cline{5-11}
22                    &                               &                                         &                           & \xmark       & \multirow{3}{*}{\cmark} & 62.50 & 27.26 & 18.41 & 8.04                 & 26.55  \\
23                    &                               &                                         &                           & SBE                         &                                        & \textbf{59.83} & 27.16 & 16.80 & 7.91                 & \textbf{25.60$^\ddag$}  \\
24                    &                               &                                         &                           & SBE+TBE                     &                                        & 60.35 & 27.11 & 17.19 & 7.95                 & \textbf{25.79$^\ddag$}  \\
\hline\hline
25                    & \multirow{3}{*}{\tabincell{c}{CTC\\(26M)}}         & \multirow{3}{*}{\tabincell{c}{DYS\\+\\CTL}}               & \multirow{3}{*}{130.1}    & \xmark       & \multirow{3}{*}{\xmark} & 78.25 & 53.18 & 46.56 & 34.09                & 50.79  \\
26                    &                               &                                         &                           & SBE                         &                                        & \textbf{68.16} & 47.08 & 40.68 & 32.40                & \textbf{45.37$^\dag$}  \\
27                    &                               &                                         &                           & SBE+TBE                     &                                        & 68.32 & 47.81 & 39.70 & 32.75                & \textbf{45.52$^\dag$}  \\
\hline\hline
\multicolumn{1}{c}{}  & \multicolumn{1}{c}{}          & \multicolumn{1}{c}{}                    & \multicolumn{1}{c}{}      & \multicolumn{1}{c}{}        & \multicolumn{1}{c}{}                   &       &       &       & \multicolumn{1}{c}{} &
 \end{tabular}}
 \vspace{-5mm}
\end{table}

Table \ref{tab:recog-result} shows the WER performance comparison between our proposed spectro-temporal deep feature based speaker adaptation, i-Vector based adaptation and LHUC based adaptation. We follow \cite{geng2020investigation} for data augmentation. Sys.1-10 are without data augmentation, Sys.11-17 are with partial data augmentation of perturbing dysarthic speech only and Sys. 18-27 are with full data augmentation of perturbing both dysarthric and control speech. Several trends can be observed\footnote{A matched pairs sentence-segment word error based statistical significance test was performed at a significance level $\alpha=0.05$.}:

\textbf{1)} Our proposed spectro-temporal deep feature adapted systems consistently outperform the comparable baseline SI systems (Sys.4-5 \textit{vs} Sys.1, Sys.13-14 \textit{vs} Sys.11 and Sys.20-21 \textit{vs} Sys.18) by up to 3.62\% (Sys.5 \textit{vs} Sys.1) absolute (11.5\% relative) WER reduction. 
\textbf{2)} Our proposed spectro-temporal deep feature adapted systems consistently outperform the comparable baseline i-Vector adaptation (Sys.4-5 \textit{vs} Sys.2, Sys.13-14 \textit{vs} Sys.12 and Sys.20-21 \textit{vs} Sys.19) by up to and 2.63\% (Sys.5 \textit{vs} Sys.2) absolute (8.6\% relative) WER reduction.
\textbf{3)} When further compounded with LHUC adaptation, our proposed spectro-temporal deep feature adapted systems consistently outperform the comparable baseline LHUC adaptation (Sys.9-10 \textit{vs} Sys.8, Sys.16-17 \textit{vs} Sys.15 and Sys.23-24 \textit{vs} Sys.22) by up to 1.39\% (Sys.10 \textit{vs} Sys.8) absolute (4.9 \% relative) WER reduction.
\textbf{4)} Temporal basis embedding features contribute to adaptation of speakers with low intelligibility (Sys.10,14,17,21) especially when the amount of training data is small. 
\textbf{5)} Experiments on the CTC E2E system (Sys.25-27) suggests our proposed spectro-temporal deep feature adapted systems also reduced the WER by 5.42\% (Sys.26 \textit{vs} Sys.25) absolute (10.67\% relative) over the comparable baseline unadapted system. 

A comparison between previously published ASR systems on UASpeech and our system is shown in Table \ref{tab:compare}.

\begin{table}[ht]
  \vspace{-2mm}
  \setlength{\abovecaptionskip}{0.05cm}
  \caption{A comparison between published systems on UASpeech and our system. Here ``DA'' refers to data augmentation.}
  \label{tab:compare}
  \centering
  \renewcommand\arraystretch{1}
  \scalebox{0.8}{
\begin{tabular}{cc}
\toprule
    Systems   &  WER\%   \\
\midrule
    Sheffield-2013 Cross domain augmentation \cite{christensen2013combining}    & 37.50 \\
    Sheffield-2015 Speaker adaptive training  \cite{sehgal2015model}    & 34.80  \\
    CUHK-2018 DNN System Combination \cite{yu2018development}         & 30.60  \\
    Sheffield-2020 Fine-tuning CNN-TDNN speaker adaptation \cite{xiong2020source} & 30.76 \\
    CUHK-2020 DNN + DA + LHUC SAT \cite{geng2020investigation}                              & 26.37 \\ 
    CUHK-2021 LAS + CTC + Meta-learning + SAT \cite{wang2021improved}
                        & 35.00 \\
    CUHK-2021 QuartzNet + CTC + Meta-learning + SAT \cite{wang2021improved}
                        & 30.50 \\
    \textbf{DA + SBE Adapt + LHUC SAT (Table 2, Sys.23)}  & \textbf{25.60} \\ 

\bottomrule
\end{tabular}}
\vspace{-5mm}
\end{table}

\section{Conclusions}
This paper presents novel spectro-temporal deep features for both accurate speech intelligibility assessment and auxiliary feature based speaker adaptation. Experiments conduct on UASpeech suggest that our proposed spectro-temporal deep feature adaptation consistently outperform i-Vector adaptation. Future research will focus on improving spectral and temporal basis features extraction for variable length speech data and application to more advanced neural network systems.

\section{Acknowledgement}

This research is supported by Hong Kong RGC GRF grant No. 14200218, 14200220, TRS T45-407/19N, ITF grant No. ITS/254/19, and SHIAE grant No. MMT-p1-19.

\bibliographystyle{IEEEtran}
\bibliography{mybib}

\end{document}